\documentclass[apjl]{emulateapj}
\usepackage{color}

\usepackage{graphicx}
\usepackage[normalem]{ulem}

\usepackage[colorlinks=true, allcolors=blue]{hyperref}

\begin{document}

\title{\textit{JWST}/NIRC\lowercase{am} Imaging of Young Stellar Objects. I. Constraints on Planets Exterior to The Spiral Disk Around MWC~758}

\color{black}
\shorttitle{Imaging MWC~758 with \textit{JWST}/NIRCam}
\shortauthors{Wagner et al.}
\author{Kevin Wagner,\altaffilmark{1}$^{\dagger \star}$ Jarron Leisenring,\altaffilmark{1} Gabriele Cugno,\altaffilmark{2} Camryn Mullin,\altaffilmark{3} Ruobing Dong,\altaffilmark{3} Schuyler G. Wolff,\altaffilmark{1} Thomas Greene,\altaffilmark{4} Doug Johnstone,\altaffilmark{5,3} Michael R. Meyer,\altaffilmark{2} Charles Beichman,\altaffilmark{6} Martha Boyer,\altaffilmark{7} Scott Horner,\altaffilmark{4} Klaus Hodapp,\altaffilmark{8} Doug Kelly,\altaffilmark{1} Don McCarthy,\altaffilmark{1} Tom Roellig,\altaffilmark{4} George Rieke,\altaffilmark{1} Marcia Rieke,\altaffilmark{1} Michael Sitko,\altaffilmark{9} John Stansberry,\altaffilmark{7} \& Erick Young\altaffilmark{10}}


\altaffiltext{1}{Department of Astronomy and Steward Observatory, University of Arizona, USA}
\altaffiltext{2}{Department of Astronomy, University of Michigan, USA}
\altaffiltext{3}{Department of Physics and Astronomy, University of Victoria, Victoria, BC, V8P 5C2, Canada}
\altaffiltext{4}{NASA Ames Research Center - Ames Center for Exoplanet Studies (ACES), CA, USA}
\altaffiltext{5}{NRC Herzberg Astronomy and Astrophysics, 5071 West Saanich Rd, Victoria, BC, V9E 2E7, Canada }
\altaffiltext{6}{NASA Exoplanet Science Institute, Jet Propulsion Laboratory, California Institute of Technology, USA}
\altaffiltext{7}{Space Telescope Science Institute, Baltimore, MD, USA}
\altaffiltext{8}{Institute for Astronomy, University of Hawaii, USA}
\altaffiltext{9}{Space Science Institute, CO, USA}
\altaffiltext{10}{Universities Space Research Association, MD, USA}

\altaffiltext{$\dagger$}{NASA Hubble Fellowship Program $-$ Sagan Fellow}
\altaffiltext{$\star$}{Correspondence to: kevinwagner@arizona.edu}


\begin{abstract}
~
MWC~758 is a young star hosting a spiral protoplanetary disk. The spirals are likely companion-driven, and two previously-identified candidate companions have been identified---one at the end the Southern spiral arm at $\sim$0$\farcs$6, and one interior to the gap at $\sim$0$\farcs$1. With \textit{JWST}/NIRCam, we provide new images of the disk and constraints on planets exterior to $\sim$1". We detect the two-armed spiral disk, a known background star, and a spatially resolved background galaxy, but no clear companions. The candidates that have been reported are at separations that are not probed by our data with sensitivity sufficient to detect them$-$nevertheless, these observations place new limits on companions down to $\sim$2 M$_{\rm Jup}$ at $\sim$150 au and $\sim$0.5 M$_{\rm Jup}$ at $\gtrsim$ 600 au. Owing to the unprecedented sensitivity of \textit{JWST} and youth of the target, these are among the deepest mass-detection limits yet obtained through direct imaging observations, and provide new insights into the system's dynamical nature.

~


\end{abstract}

\section{Introduction}

Protoplanetary disks host a variety of substructures (e.g., \citealt{Muto2012}), some of which are driven by forming planets (e.g., \citealt{Keppler2018, Wagner2018, Currie2022}). Giant planets create gaps and launch spiral waves that can be more readily detectable than the planets themselves. As spiral arms are launched on the dynamical timescale ($\sim 10^3$ yr for a planet with an orbital semi-major axis of $a=100$ au around a 1 $M_\odot$ star), spiral density waves should be as common as the giant planets that drive them \citep{Zhu2015}. 

Indeed, several systems have been shown to host spirals (see review in \citealt{Dong2018,Bae2022}). However, the anticipated planets themselves remain seldom detected---perhaps due to low initial temperatures (e.g., \citealt{Marley2007}) or due to attenuation by dust. In either case, such planets should be most readily observable at infrared wavelengths. For this reason, observations with the James Webb Space Telescope (\textit{JWST}) are expected to push detection limits significantly further. 

~





MWC~758 is a 3.5$\pm$2 Myr old Herbig A8Ve member of the Taurus star forming association \citep{Meeus2012}. \cite{Luhman2023} suggest a slightly older age of $\sim$18 Myr for MWC 758 based on its kinematic association with other young stars. However, the existence of the protoplanetary disk and accretion rate of the star both favor the younger ages found in the literature \citep{Ribas2015}, which we adopt here. The stellar mass is estimated to be 1.5$-$1.9 M$_\odot$ (\citealt{Vioque2018, Garufi2018}) and a distance of d=156$\pm$1 pc was measured by the \cite{GaiaDR3}. Due to its youth and presence of a disk, this system has been the focus of several high-contrast imaging studies, as well as interferometric radio observations and optical/infrared spectroscopy. 

\cite{Grady2013} first revealed the spiral arms in scattered light. From $H$-band polarized intensity and $K^\prime$ total intensity observations with Subaru/HiCIAO), they identified the two main spiral arms (hereafter the Northern and Southern spirals). MWC~758's spiral disk is seen $\sim$21$^\circ$ from face-on \citep{Isella2010} and displays a clear two-armed ($m$=2) symmetry that is characteristic of a massive external perturbing companion (i.e., a giant planet to a low-mass star: \citealt{Fung2015}). Through infrared spectroscopy, \cite{Grady2013} also revealed that the central star and inner disk display near-infrared variability of up to 20\% over timescales of a few years.



Two recent works have explored the protoplanetary disk around MWC~758 with radio interferometric observations from ALMA. \cite{Dong2018b} presented 0.87 mm continuum observations with $\sim$40 mas angular resolution $-$ corresponding to $\sim$7 au projected separation at 160 pc. These observations detected the dust disk out to a separation of 0$\farcs$64 (102 au) with SNR$>$3, revealing also the central cavity, two broad dust clumps, and three distinct ring structures. The Southern spiral arm was marginally detected above the background disk emission, indicating that it is possibly the more massive (and thus primary) spiral arm --- i.e., the one that ought to be nearer to the driving companion. The ALMA observations of \cite{Boehler2018} reveal the $^{13}$CO and C$^{18}$O gas emission, which are also both in steep decline at $0\farcs6$, or $\sim$100 au. The Southern spiral is likewise more prominent in the gas than the Northern spiral. These observations are broadly suggestive of a giant planet exterior to the Southern spiral arm. 

\cite{Cugno2019} used SPHERE/ZIMPOL to search for H$\alpha$ ($\lambda$=656 nm) emission from accreting protoplanets in MWC~758. No sources were detected, and the authors established a background-limited contrast of $\sim$10 magnitudes in the H$\alpha$ filter, after continuum subtraction. \cite{Huelamo2018} and \cite{Zurlo2020} also used SPHERE/ZIMPOL to search for H$\alpha$ emission, ultimately reaching a signal to noise (S/N)=5 contrast ratio of 9.5 magnitudes in the H$\alpha$ filter (similarly, with nearby continuum subtracted). They also did not detect any emitting sources. Likewise, the observations of \cite{Grady2013} established sensitivity to 3$-$4 M$_{\rm Jup}$ companions at 1" and 2 M$_{\rm Jup}$ at 2", respectively, assuming negligible attenuation due to circumstellar and/or cirucmplanetary dust and hot-start evolutionary models (e.g., \citealt{Baraffe2003, Marley2007}). These null results are consistent with the expected level of attenuation of accreting protoplanets, which can be tens to hundreds of magnitudes at optical wavelengths (e.g., \citealt{Szulagyi2019,Chen2022}).

Keck/NIRC2 observations of the system revealed the two well-known arms in the $L^\prime$ filter, constituting the first detection of the disk in the thermal IR \citep{Reggiani2018}. The observations were also the first to utilize angular differential imaging (ADI: \citealt{Marois2006}), which distorts the appearance of the disk features --- including spiral arms --- due to the negative side-lobes introduced adjacent to positive sources. These observations revealed two potentially new sources within MWC~758: a putative third spiral arm and a potential companion (MWC~758b) interior to the disk gap at 0$\farcs$11. However, neither feature was recovered in more sensitive observations in the same filter \citep{Wagner2019}.

The Large Binocular Telescope Interferometer (LBTI) observed MWC~758 five times over 2016$-$2019. Three observations were taken in direct imaging mode with two observations in the $L^\prime$ filter and one in $M^\prime$, which were published in \cite{Wagner2019}. Two follow-up observations were taken in the spectroscopic ALES mode (\citealt{Skemer2018,Stone2022}). The initial observations \citep{Wagner2019} revealed a candidate point source at $\sim$0$\farcs$62 at the end of the Southern spiral arm. \cite{Wagner2023} presented follow-up spectroscopy that revealed a red spectrum of the source (henceforth MWC~758c) that is distinct from the rest of the disk. MWC~758c is consistent with a mass of $\gtrsim$2-4 M$_{\rm Jup}$ --- i.e., it is consistent with driving the spiral arms. 

While most available evidence points to MWC~758c being the planet responsible for driving the spiral arms, a possibility that cannot be completely excluded is that MWC~758c is a disk feature with apparently planet-like properties (i.e., a very red spectrum, position at the end of a spiral arm, lack of polarized counterpart, etc.). Indeed, \cite{Ren2020} have suggested that the perturbing companion could instead be at a wider separation. Their argument is based on the spiral arm rotation rate, and while the best-fit rotation rate favors a driving companion with a separation of $\sim$1$\farcs$1, or $\sim$172 au, the uncertainty on the rotation rate does not rule out MWC~758c at $\sim$0$\farcs$6, or $\sim$96 au, as their driver \citep{Wagner2023}.


Now, \textit{JWST}/NIRCam enables pushing the detection limits at $\gtrsim$1" to sub-Jovian masses. A non-detection of planets or more massive companions exterior to MWC 758c would rule out alternative hypotheses for a more distant spiral arm driving planet, leaving MWC~758c as the most likely body responsible for driving the spiral arms. However, the inverse is not necessarily true: i.e., the detection of a possible second planet with NIRCam would not necessarily preclude that MWC~758c could be the planet responsible for the spiral arms. Furthermore, should another planet exist, its discovery would play an important role in helping to understand this system's formation and the dynamical evolution of young planetary systems. In a companion paper \citep{Cugno2023}, we present an analysis of the SAO 206462 system, which shows a similar spiral morphology.


\begin{figure}[htpb]
\epsscale{1.12}
\plotone{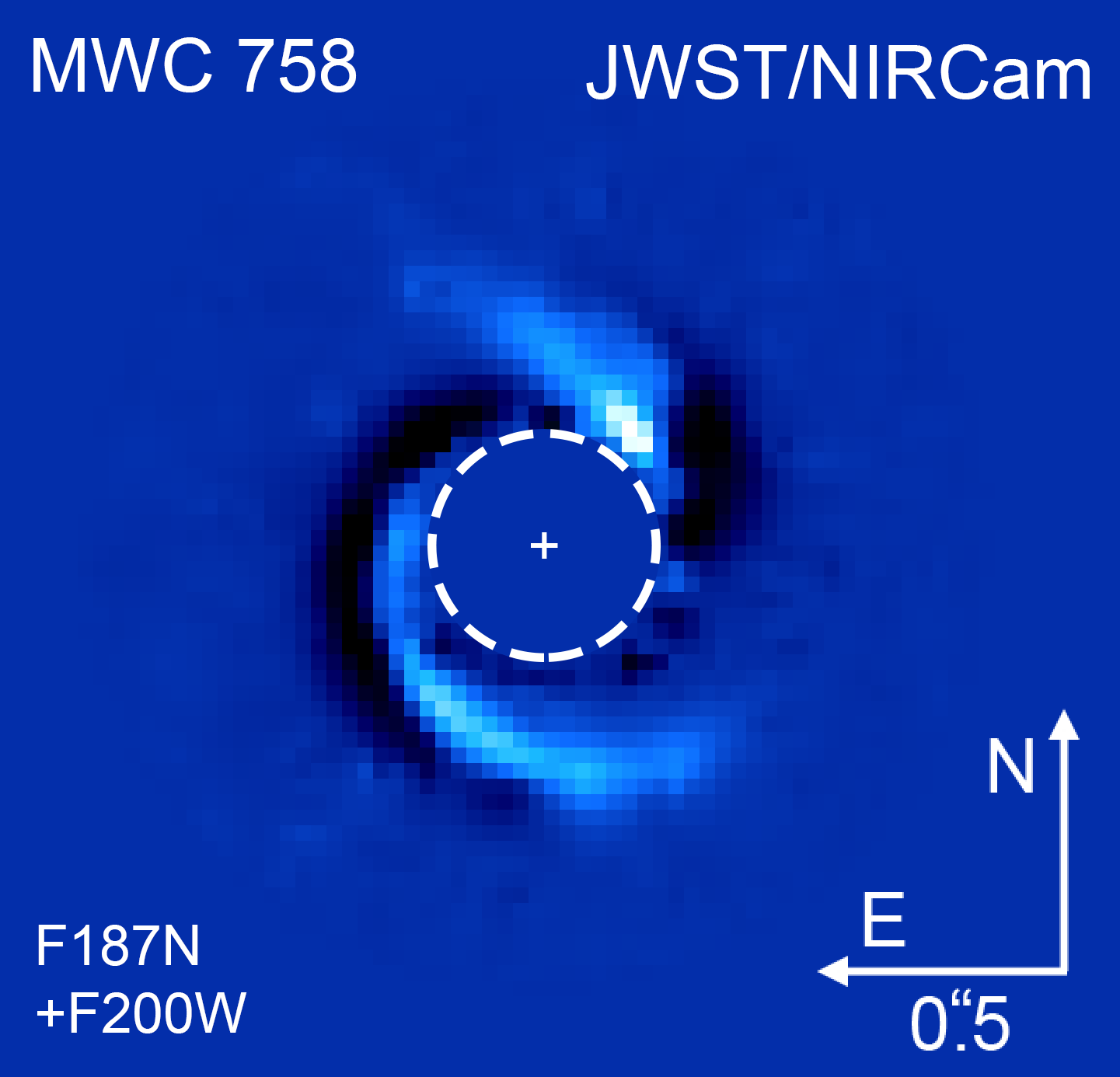}
\caption{\textit{JWST}/NIRCam image of MWC~758's protoplanetary disk in the F187N+F200W filters. No coronagraph was used, although a software mask was placed over the inner speckle-dominated region. The location of the star is marked by the crosshairs. The images were processed with angular differential imaging (specifically, ADI-KLIP: \citealt{Soummer2012}), normalized to the brightest part of the disk, and then added to one another. The two spiral arms are clearly visible. The image is the result of Roll 1 - Roll 2. Negative regions to the sides of the disk are the result of the angular differential imaging processing.}
\label{disk}
\end{figure}

\section{Observations and Data Reduction}

\label{observe}


We observed MWC~758 with \textit{JWST}/NIRCam \citep{Rieke2023} as part of the guaranteed time observations (GTO) to directly image Young Stellar Objects (YSOs; PID 1179; PI Leisenring). Data were collected on UT 2022-09-30 in the direct (non-coronagraphic) imaging mode in the F187N, F200W, F405N, and F410M filters of NIRCam. The narrowband filters (F187N and F405N) are centered on hydrogen emission lines, whereas the F200W and F410M filters capture more continuum emission from expected protoplanet spectra. In each filter, we obtained observations at two spacecraft roll orientations (separated by 10$^\circ$) to enable angular differential imaging. The NIRCam detector was operated in the RAPID readout pattern with 10 groups per integration, in which all 10 read frames up-the-ramp are stored individually in the final datacube. We acquired data in the SUB160P subarray with corresponding frame times of 0.27864sec.\footnote{This setup was chosen in order to provide the deepest sensitivity outside of 1". Deeper observations interior to 1" for similarly bright targets could be obtained with either the SUB64P array, or with a coronagraph.} The non-destructive ramp sampling enables identification of pixels that are saturated, non-linear, or affected by cosmic rays throughout the readout sequence of a given integration. However, many pixels saturate within the first group of our observations. To mitigate the effects on the measured signal levels of charge spillage from neighboring saturated pixels, it is also beneficial to set a maximum $N_{group}$ based on the flux of the surrounding pixels (see below). We collected 480 integrations per roll position and filter pair, for a total of 960 total integrations per filter and maximum exposure time of 2675sec ($\approx$3/4 hr). We used a four point subarray dither pattern in order to mitigate the impact of detector artifacts.

We reduced the data using a custom-built set of IDL reduction scripts specifically designed to handle this dataset (in particular the large number of pixels affected by charge transfer and centering of saturated data). Our data reduction process is described in more detail below. 





We began with the raw \texttt{\_uncal.fits} files, which are the original data products saved by NIRCam. First, we subtracted the superbias images from the \textit{JWST} Calibration Reference Data System.\footnote{Specifically, we used the NRCBLONG INFLIGHT 2022-04-14 2022-08-05 and NRCB1 INFLIGHT 2022-07-28 2022-08-05 calibrations from \url{https://jwst-crds.stsci.edu}.} Next, for the long wavelength (LW) subarrays we subtracted the mean of the reference pixels at the edge of the image, frame-by-frame. We then applied a linearity correction following \cite{Canipe2017}. We converted each integration to a slope image (units of DN/sec) by fitting the level of charge vs. group number. In order to mitigate the effects of saturation, cosmic rays, and charge transfer, the maximum group number can be truncated, which effectively shortens the exposure time. We limited the number of groups to [4,3,2] for pixels whose maximum of adjacent surrounding rates was greater than [500,1000,2000] counts/group (fit to the total number of groups). 

We identified cosmic rays as pixels with a rate of at least 100 counts/group (fit to the total number of groups) and in which the counts in any single group were at least 10$\times$ greater than the counts in the previous group. For these pixels, we truncated the maximum number of groups at the one preceding the jump in counts. Finally, we scanned each slope image individually for bad pixels with a sigma clipping routine (with thresholds of 6,6,5,4 sigma for F187N, F200W, F405N, and F410M) in boxes of $6\times6$ surrounding pixels (except for F410M, for which we used $10\times10$ pixels). These values were chosen by eye to account for a majority of bad pixels. Other obviously bad pixels (those appearing as single bright pixels in every slope image) were added manually to the mask. This typically resulted in $\sim$2\% bad pixels per frame, which were replaced with a median of the surrounding pixels. 


Next, we super-sampled the slope images by a factor of four in order to mitigate errors due to interpolation in the image alignment and centering procedures. We aligned the images via cross correlation with the first image in the sequence. The standard deviation of image shifts were $\sim$1.1---1.5 mas, which is consistent with telescope pointing jitter (expected to be $<$2 mas). We then found the center via a modification of the rotational centering algorithm presented in \cite{Morzinski2015}, which is typically accurate to $\lesssim$0.25 pixels (we verified a similar level of accuracy of $\lesssim$0.2 pixels for NIRCam using simulated images). In short, this method utilizes the rotational symmetry of the PSF in order to determine the precise center. This approach is justified when the central regions are affected by saturation, causing  conventional centroid approaches or PSF-fitting approaches to fail. For each tested position, the algorithm rotates the image cube by 180$^\circ$ and subtracts the median of the cube from each image. The residual cube is then median combined, and the position resulting in the least squared residuals is taken as the image center. We used a grid of 2$\times$2 pixels with a step size of 0.05 pixels. We limited the image area to the 40$\times$40 pixel region surrounding the estimated center of the saturated region and excluded pixels within a radius of 8 pixels from the estimated center (12 for F410M data). We then resampled the images to their native resolution. 

We rejected integrations with a maximum cross correlation function less than 0.9992, 0.99993, 0.9993, and 0.99997 (for F187N, F200W, F405N, and F410M, respectively) with respect to a rolling median of the surrounding twenty frames, which corresponds to 2\%, 8\%, 0.3\%, and 6\% of frames (for F187N, F200W, F405N, and F410M, respectively). These values were iteratively tested and chosen as those that provided the deepest average sensitivity curves based on synthetic point source injections (see \S \ref{results}). The final results were not very sensitive to frame rejection percentages below these values (up to and including using all collected frames). 

At this stage, where relevant, we injected synthetic point sources with PSFs for each filter generated using {\tt WebbPSF} \citep{Perrin2014, Leisenring2022}. Since MWC~758 is saturated in the images, in order to convert from sensitivity to contrast we used the spectrum from IRTF/SpeX taken on 2021-02-03 and presented in \cite{Wagner2023}, which can be seen in Appendix \ref{Appendix A}. MWC 758 has an intrinsic variability of $\lesssim$20\% \citep{Grady2013}, which should be considered when comparing these results to ground-based data, which are typically reported in terms of contrast to the host source without absolute photometric calibrations.

\begin{figure*}[htpb]
\epsscale{1.12}
\plotone{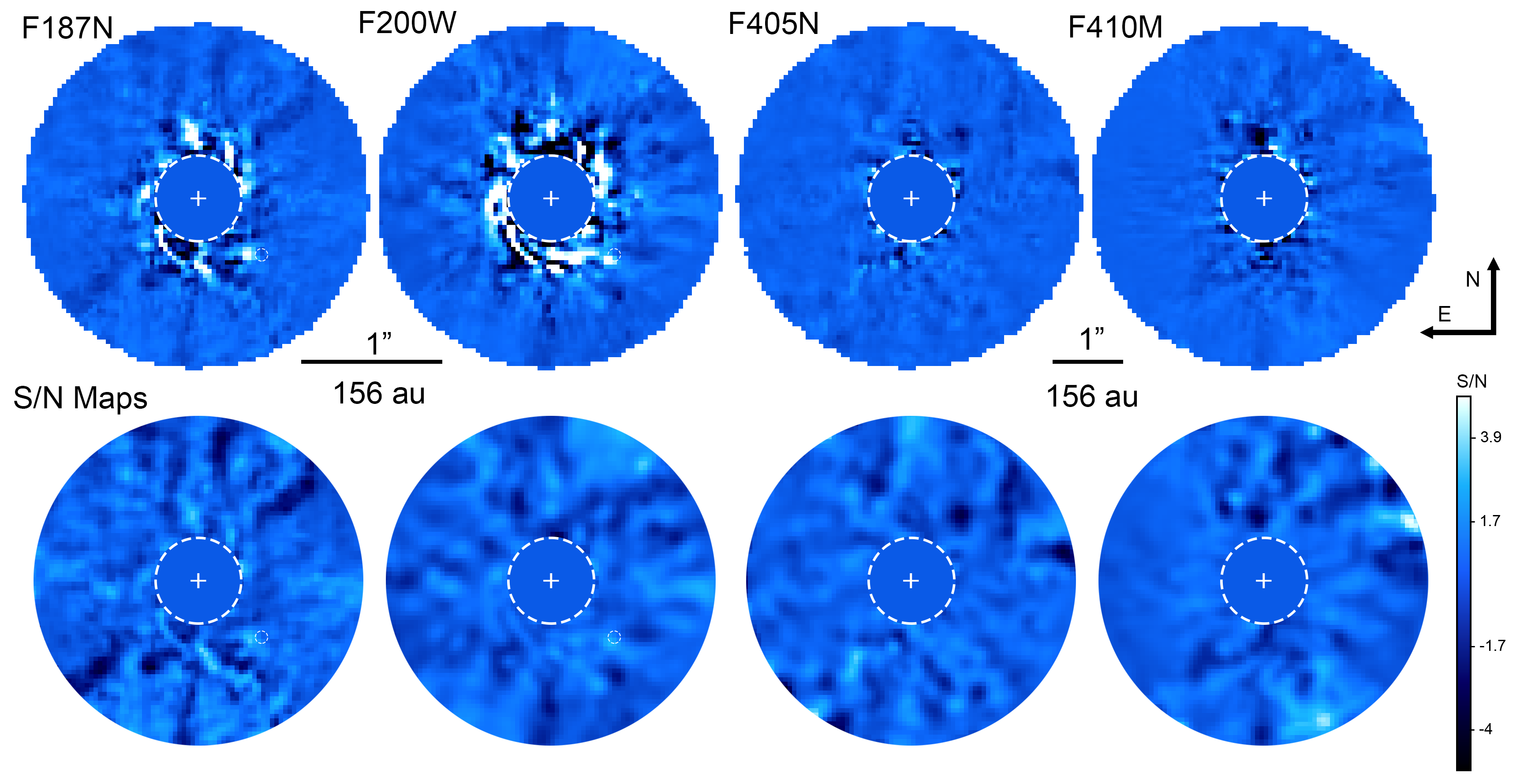}
\caption{\textit{Top:} \textit{JWST}/NIRCam images of MWC~758 processed with ADI-KLIP. \textit{Bottom:} Corresponding signal to noise ratio (S/N) maps. Note that the long$-$ and short-wavelength images have different platescales, and that the software mask is twice as large for the long-wavelength images. The location of MWC 758c \citep{Wagner2023} is indicated in the short-wavelength images (it is at the edge of the mask in the long-wavelength images). There is a low-S/N ($\sim$3) source in MWC 758c's vicinity in the F187N and F200W images. Aside from the spiral protoplanetary disk, no other objects are detected with S/N$\geq$5 within separations of 2". At wider separations, two background objects are detected (see Appendix \ref{Appendix B} \& \ref{Appendix C}).}
\label{klip}
\end{figure*}

\begin{figure*}[htpb]
\epsscale{1.17}
\plotone{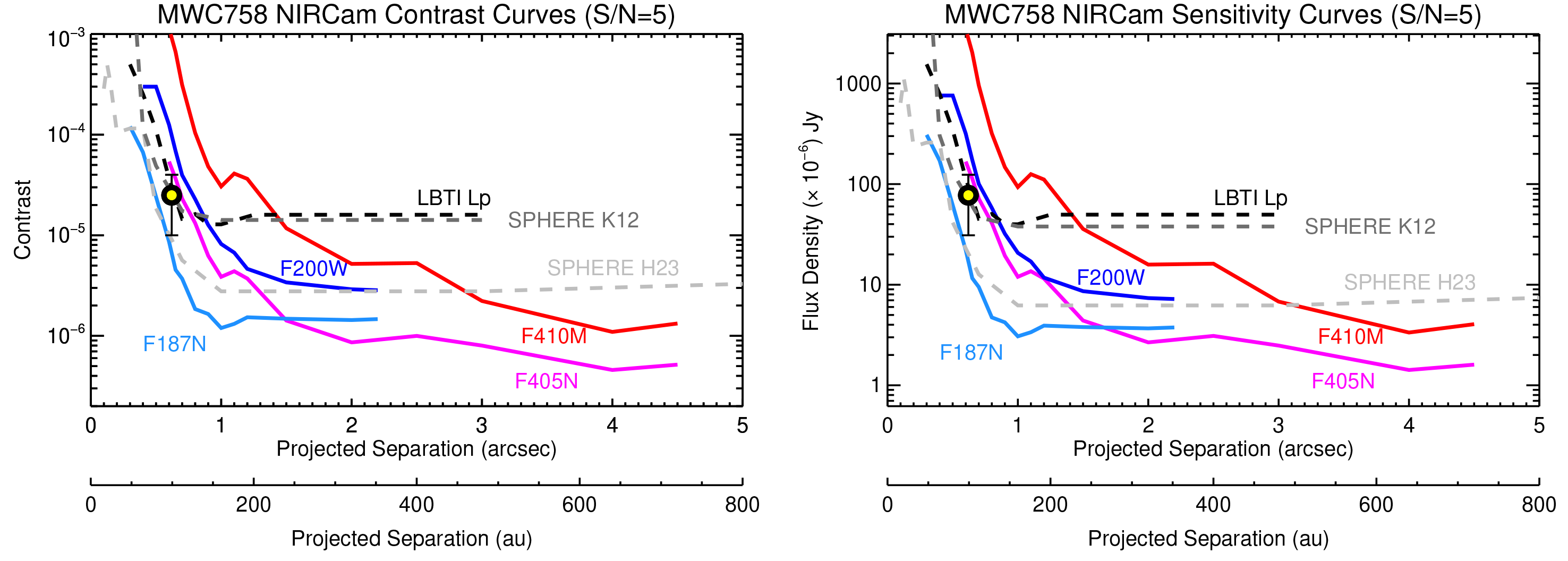}
\caption{Contrast curves generated via synthetic point source injections. Comparisons to ground-based data \citep{Wagner2019,Boccaletti2021} at similar wavelengths are shown in dashed curves. F187N is most comparable in wavelength to SPHERE's $H23$ filter, F200W is most comparable to SPHERE's $K12$ filter, and F405N and F410M are most comparable to LBTI's $Lp$ (or $L^\prime$) filter. The yellow circle corresponds the brightness of MWC 758c measured at 4.05 $\mu$m with LBTI/ALES \citep{Wagner2023}. At wavelengths longer than $\lambda \gtrsim$ 2 $\mu$m and projected separations $\gtrsim$1", the NIRCam data reach over an order of magnitude fainter sensitivities than ground-based data. These sensitivities are converted to mass detection limits in Fig \ref{mass}. The scattered light disk extends to $\sim$0$\farcs$55 \citep{Benisty2015}. Sensitivities interior to this are likely under-estimated as a result. }
\label{cont}
\end{figure*}

\begin{figure*}[htpb]
\epsscale{1.17}
\plotone{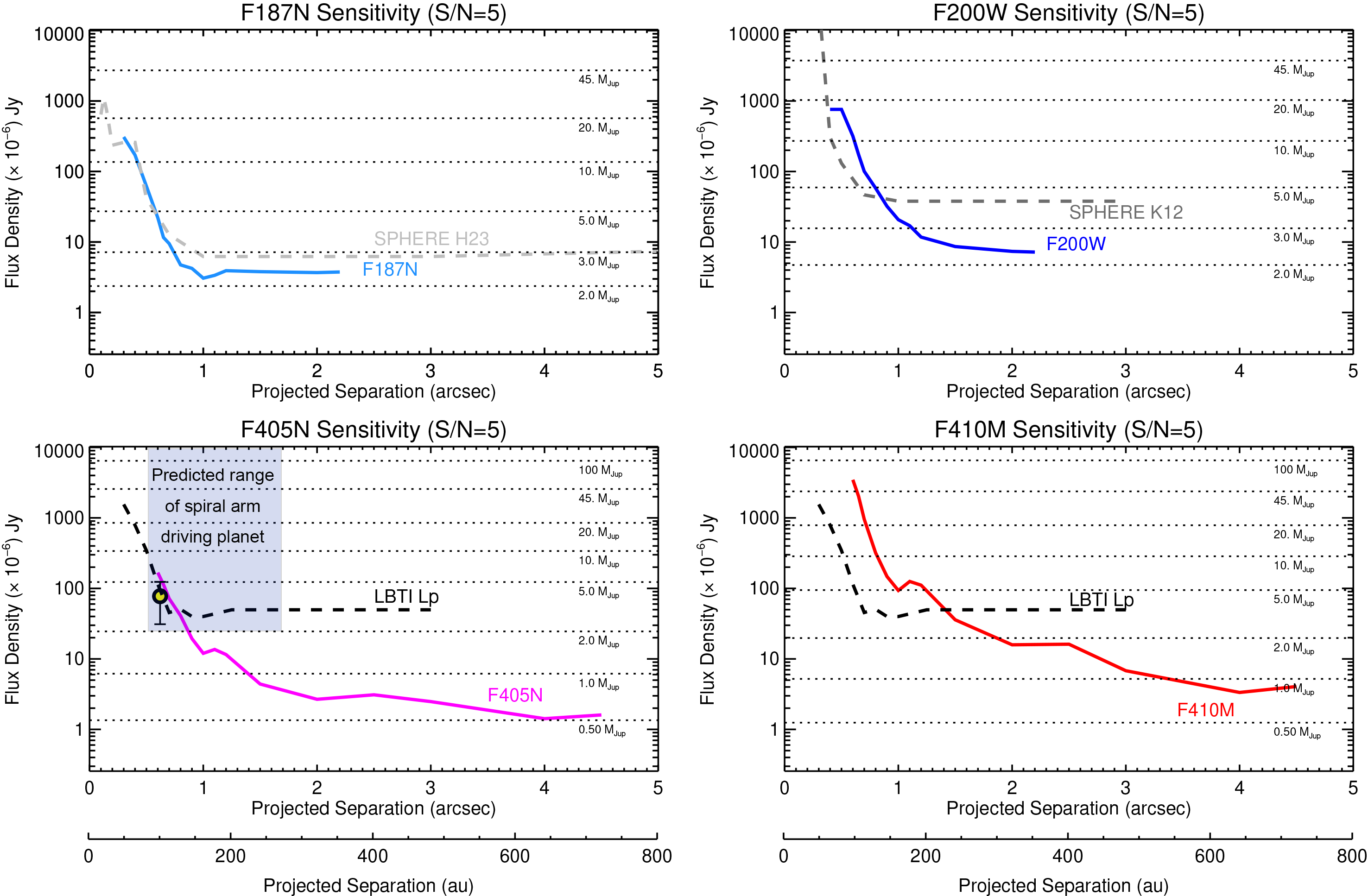}
\caption{Mass sensitivity for each filter generated via synthetic point source injections. Compared to ground-based data (dashed curves), at separations $\gtrsim$1" the NIRCam data reach lower masses in all filters. In the background limit, the F405M and F410M filters reach masses that are $\sim$5 times lower than the LBTI data \citep{Wagner2019}. The F405N data provide exceptionally deep sensitivity limits of $\gtrsim$0.5 M$_{\rm Jup}$, owing to the brightness of young planets at $\sim$4$\mu$m and \textit{JWST}'s far superior background-limited sensitivity. The yellow circle corresponds the brightness of MWC 758c measured at 4.05 $\mu$m with LBTI/ALES \citep{Wagner2023}. The mass of this object may be higher than $\sim$5 M$_{\rm Jup}$ for higher levels of extinction. The minimum and maximum range of predicted parameters for the spiral arm driving planet are shown in the gray shaded region. The range of plausible combinations of these parameters has a more complex shape $-$ i.e., more widely separated companions require a greater mass to drive similar spiral arms: see \cite{Dong2015}, \cite{Fung2015}, and \cite{Ren2020} further details.}
\label{mass}
\end{figure*}

\begin{figure*}[htpb]
\epsscale{1.17}
\plottwo{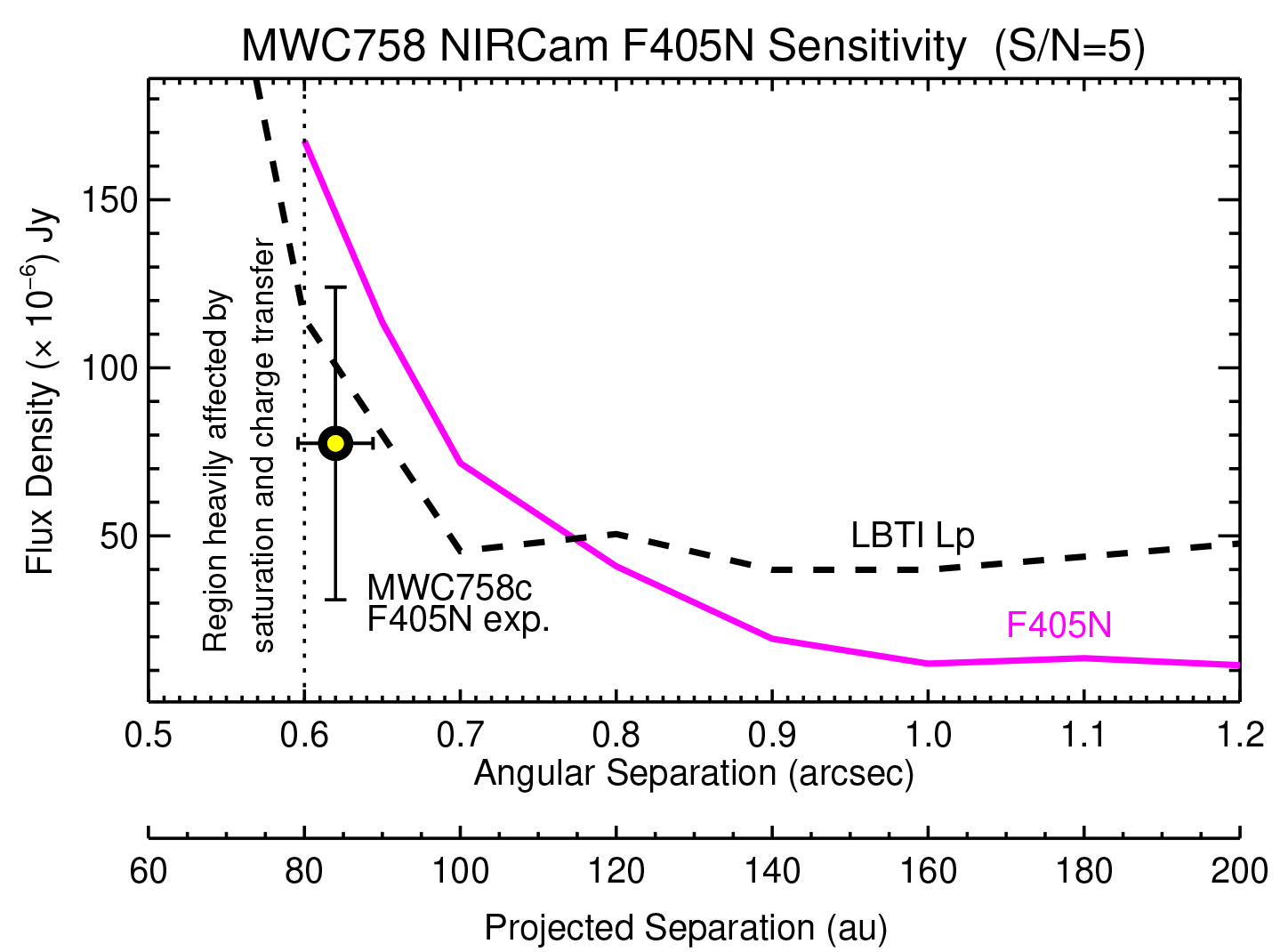}{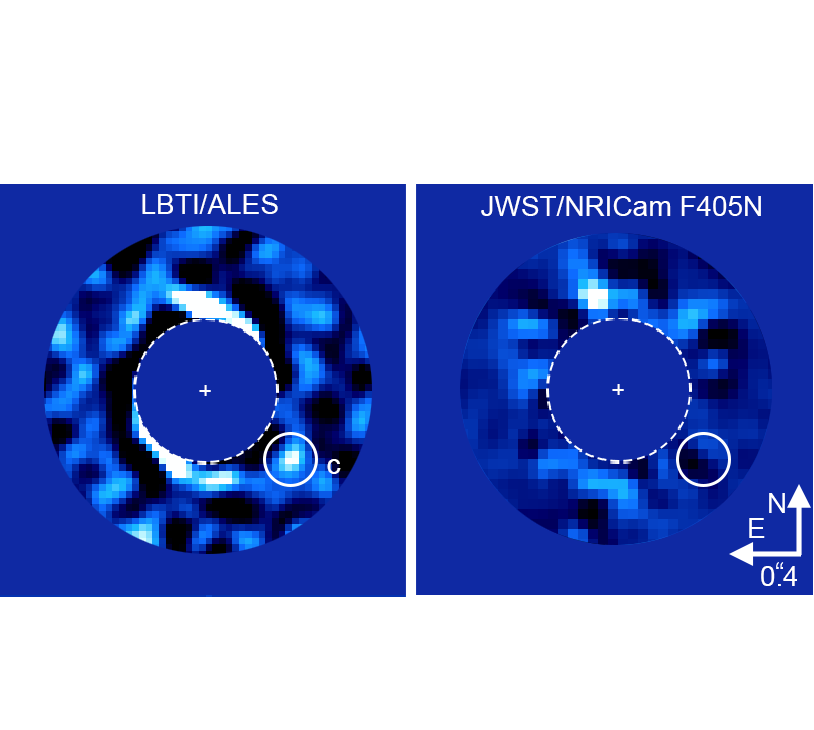}
\caption{\textit{Left: }Contrast curves generated from synthetic point source injections (also shown as mass detection limits in Fig \ref{mass}). \textit{Center: }LBTI/ALES 3.97-4.14 $\mu$m image, and \textit{right:} \textit{JWST}/NIRCam F405N data with a smaller optimization range (0$\farcs$25$-$1$\farcs$2) focused on recovering closer-in sources. The yellow circle corresponds the brightness of MWC 758c measured at 4.05 $\mu$m with LBTI/ALES \citep{Wagner2023}. The F405N data are not sensitive enough to detect MWC~758c, which will likely require the use of a coronagraph. The spiral arms are weakly detected in the F405N data.}
\label{MWC758c}
\end{figure*}

We binned the data by 20 integrations using a median and then subtracted the PSF via two independent algorithms: the first via a simple application of ADI \citep{Marois2006}, which can be seen in Appendix \ref{Appendix B}. The second PSF subtraction algorithm used projection onto eigenimages (Karhunen-Lo\`eve image processing, or KLIP: \citealt{Soummer2012}), which is similar to the principal component analysis method (PCA: \citealt{Amara2012}). Specifically, we used the KLIP implementation in \cite{Apai2016}, with $K_{KLIP}$=4 (equivalent to four principal components) over annular segments of 10 pixels in radius and 45$^\circ$ in azimuth, and using all images from the opposite roll as the reference set.

We generated contrast curves via injecting point sources as described above and iterating upon the brightness until it was within 10\% of the chosen threshold (using S/N=5 as the threshold and calculating S/N according to \citealt{Mawet2014}). For shortwave (SW) data, we tested separations of [0.3, 0.4, 0.5, 0.6, 0.65, 0.7, 0.8, 0.9, 1.0, 1.1, 1.2, 1.5, 2.0, 2.2] arcsec and for longwave (LW) data we tested separations of [0.6, 0.65, 0.7, 0.8, 0.9, 1.0, 1.1, 1.2, 1.5, 2.0, 2.5, 3.0, 4.0, 4.5] arcsec. For both SW and LW data, we tested position angles spaced uniformly 10$^\circ$ apart, beginning at 0$^\circ$. For all filters, we measured the signal and noise in all non-overlapping apertures of one FWHM diameter. We combined results at each separation by taking a median over the sensitivities obtained at different position angles.

~
~

\section{Results}

\label{results}

The KLIP-processed images and S/N maps are shown in Figs.\ \ref{disk}, \ref{klip}. The disk itself is heavily distorted by the ADI-style image processing combined with the spatially extended nature of the nearly face-on disk and the small amount of field rotation (10$^\circ$). The individual roll subtraction images (i.e., combinations of individual halfs of the total observing sequence, rather than the entire derotated cube) yield cleaner results on the disk (see Fig. \ref{disk}). However, negative over-subtraction regions adjacent to the brighter parts of the disk in the direction of the spacecraft roll are plainly visible. Appendix \ref{Appendix B} shows the classical-ADI processed versions of each individual roll. 

For point sources, the combination of the two rolls provides the best sensitivity. These images are shown in Fig.~\ref{klip} along with their corresponding S/N maps. Despite some structure in the images (mostly attributable to residuals from the disk after undergoing ADI-processing), no source is detected with S/N$>$3. One source appears close with S/N=2.9 in the F187N image. This source is also within $\sim$0$\farcs$1 of the position of MWC~758c and will be discussed in \S \ref{discuss}. At separations wider than 2", the known background star \citep{Grady2013} is detected to the Northwest of MWC~758. This source is detected with such S/N that it needs to be masked in the KLIP reductions that extend out to its separation in order to not bias the subtraction of the PSF of MWC~758. An elongated source is also detected to the Northeast of MWC~758 whose colors and morphology are consistent with a background galaxy (see Appendix \ref{Appendix C}).


As described in \S \ref{observe}, we generated senstivity curves via synthetic point source injection and retrievals. These are shown in Fig. \ref{cont}. We also compiled data from ground-based observations from \cite{Wagner2019,Wagner2023} and \cite{Boccaletti2021}. The NIRCam sensitivity exceeds that of the ground-based observations in comparable filters at separations greater than 0$\farcs$6, 0$\farcs$9, 0$\farcs$8, and 1$\farcs$5 for F187N, F200W, F405N, and F410M, respectively. In the F405N and F410M filters, the background-limited sensitivity at $\sim$4" exceeds that achieved by ground-based observations by more than a factor of ten.

The sensitivity limits were converted to mass detection limits via the BEX evolutionary models of \cite{Linder2019} for up to 2 M$_{\rm Jup}$ with a smooth interpolation to the AMES-COND models for higher masses \citep{Baraffe2003}. We assumed an age of 3.5 Myr and negligible attenuation due to dust. The lack of dust is likely a reasonable assumption for separations $\gtrsim$1", as this is sufficiently removed from the circumstellar material. However, circumplanetary material would raise the mass detection limit for a given sensitivity (e.g., \citealt{Szulagyi2019}). The deepest limits are reached in the F405N filter, which is sensitive to $\sim$0.5 M$_{\rm Jup}$ planets. At separations of $\sim$1", the data are sensitive to planets of 2 M$_{\rm Jup}$, and at separations of $>$1$\farcs$0 planets of 1 M$_{\rm Jup}$ could be detected. At the separation of MWC~758c, the sensitivity of the F405N data approaches that of the LBTI observations \citep{Wagner2019,Wagner2023}, but ultimately is a factor of $\sim$2$-$3 too shallow in order to detect MWC~758c, as shown in in Fig. \ref{MWC758c}. A further caveat to note that arises from the system's young age is the assumption of post-formation evolutionary models; whereas we may actually be observing planets in the process of formation. However, the direction in which this would bias our results is unclear, as planetary luminosities during formation are not well constrained. Note also that recent work has suggested that MWC 758 may belong to a slightly older association \citep{Luhman2023}. If that is the case then the evolutionary models (with an assumed age of 3.5 Myr) would be over-estimating the planet brightness.

\section{Discussion}
\label{discuss}

To recap, our primary goal is to image planets and to establish mass-detection limits in the outer regions of the MWC~758 system (beyond $\sim$1", or $\sim$150 au). Since these are also among the first images taken of a protoplanetary disk-hosting system with \textit{JWST}, we also aim to image the interior regions in order to demonstrate the capability of \textit{JWST}/NIRCam to observe protoplanetary disks in parallel to searches for exoplanets. We note that these images are taken in ``direct" mode$-$i.e., without the use of a coronagraph. Results may differ for other observing modes, which we discuss at the end of this section.

The disk is detected clearly in the F187N and F200W filters, and tentatively in the F405N filter. In all images, the spiral arms are significantly impacted by over-subtraction of the bright/faint regions of the nearly face-on and spatially resolved disk. We also attempted a reference differential imaging strategy using data taken from another source within this program (SAO 206462; see \citealt{Cugno2023}); however, this object also contains a spiral disk, and the resulting detection limits for both spatially extended emission and point sources were lower than the roll-subtractions presented here. 

Numerical simulations have shown that a giant planet outside of the two spirals can reproduce their observed features in scattered light as well as the outer disk ring and vortex in mm continuum emission \citep{Dong2015, Baruteau2019}. The mass of such an outer planet may be constrained based on the symmetry of its spirals \citep{Fung2015}. For nearly symmetric pairs of arms, such as the ones in MWC~758, this method can only provide a lower limit of $\sim$4--8 $M_{\rm J}$, because companions more massive drive similarly symmetric arms \citep{Dong2016}. An even more massive companion is also expected to be further away from the disk as it opens a wider gap, if the disk has reached a steady state (i.e., if the companion did not form recently). 

While several lines of evidence point to MWC~758c at $\sim$0$\farcs$6 as being a giant planet, and therefore responsible for the spiral arms \citep{Wagner2019,Wagner2023}, further support for a driving planet at $<$0$\farcs$8 comes from the lack of detection of additional planet candidates in the NIRCam data. The F405N data is sensitive to planets that are much less massive than the mass of the planet that is predicted to be driving the spiral arms (\citealt{Fung2015}; see Fig. \ref{mass}). Therefore, the non-detection of additional planet candidates in the NIRCam data is consistent with MWC~758c being responsible for the spirals. 

This dataset is less sensitive at $\lesssim$0$\farcs$8 than the ground-based data taken with the LBT \citep{Wagner2019,Wagner2023}. This is demonstrated in Fig. \ref{MWC758c}, which shows a side-by-side comparison one of the LBT images with the F405N image from this work, along with the associated contrast curves. The F405N data are a factor of $\sim$2$-$3 too shallow (in terms of sensitivity) to have detected MWC~758c. The source's position is coincident with a negative speckle in the F405N image, likely due to imperfectly subtracted speckles among \textit{JWST}'s diffraction spikes. Coronagraphic imaging of this target in Cycle 2 (GO-04014) will likely improve these limits. In addition, we can further improve the sensitivity of this data with more advanced data reduction techniques that accurately predict and remove the substantial charge migration from saturated regions, but such an effort is beyond the scope of this paper. Finally, there is some positive flux in the vicinity of MWC~758c in the F187N and F200W filters (within $\sim$0$\farcs$1). The significance of this feature is weak (S/N$\lesssim$3); however, this low value is likely impacted by scattered light from the disk that is present within the noise apertures at the same radius. It is not clear whether this apparent source is a noise feature, or if it could possibly be associated with MWC 758c.


Finally, compared to coronagraphic observations (e.g., \citealt{Carter2023}), where F444W is typically used in order to provide the widest bandpass for deep planet searches, direct imaging of significantly brighter sources (such as MWC 758, which is $\sim$2 magnitudes brighter than HIP 65426 at $\sim$4 $\mu$m) requires narrower filters in order to mitigate issues from heavy saturation (e.g., \citealt{Argyiou2023}). Our deepest observations in the F405N filter reached $\sim$1.3 $\mu$Jy at 4 arcsec after 2675 sec of exposure. Comparatively, \cite{Carter2023} achieved a background limit of $\sim$0.7 $\mu$Jy in the F444W filter with MASK335R in 1235 sec, whereas a similar exposure time of 2675 sec would result in $\sim$0.5 $\mu$Jy. This highlights an important tradeoff between observing modes: direct imaging, without the loss of throughput introduced by the Lyot stop, can achieve similar background-limited performance in narrow bands to coronagraphic observations in wide bands. Beyond this basic comparison, there are a number of other considerations that future programs should take into account when planning observations, such as the presence of circumstellar material that may be distorted by the coronagraph, as well as varying observation configurations (such as subarray size and readout pattern) that lead to different levels of read noise and saturation.

\section{Summary and Conclusions}

These data provide one of the deepest looks yet available into a protoplanetary disk-hosting system and showcase the potential of \textit{JWST} to deliver sensitive constraints on forming planets. In summary of our results: 
\begin{enumerate}
\item We placed new limits on planets exterior to the known spiral disk, down to $\sim$0.5 M$_{\rm Jup}$ in the background limit at $\gtrsim$600 au projected separation and below 2 M$_{\rm Jup}$ at $\gtrsim$150 au (both in the F405N filter, in which we reach the deepest sensitivity). 
\item We detected the spiral disk in the F187N, F200W, and F405N filters, with some distortion from the roll-subtraction strategy. 
\item We detected the known background star to the NW of MWC 758, and an elongated object to the North that we identify as a likely background galaxy. 
\item We interpret the non-detection of additional companions as consistent with a spiral arm driving companion at the end of the Southern spiral arm at $\sim$0$\farcs$6, as identified in \cite{Wagner2019, Wagner2023}.
\end{enumerate}

~

Overall, the lack of giant planets beyond 1" in the MWC 758 system strengthens the conclusion that MWC 758c at $\sim$96 au is the body responsible for driving the spiral arms. The non-detection of additional companions at wider separations is consistent with expectations, as few wide-orbit ($a\sim$100 au) companions have been identified (e.g., \citealt{Stone2018,Nielsen2019,Vigan2021,Wagner2022}). Such distant companions are also expected to be uncommon based on planet formation models (e.g., \citealt{Pollack1996,Mordasini2009,Forgan2018}). Finally, these observations demonstrate the capabilities of \textit{JWST}'s direct imaging mode with NIRCam for observing protoplanetary disks and planets on wide orbits.

\section{Acknowledgments} The authors are grateful for support from NASA through the \textit{JWST}/NIRCam
project though contract number NAS5-02105 (M. Rieke, University of Arizona, PI). The results reported herein benefited from collaborations and/or information exchange within NASA's Nexus for Exoplanet System Science (NExSS) research coordination network sponsored by NASA's Science Mission Directorate. K.W. acknowledges support from NASA through the NASA Hubble Fellowship grant HST-HF2-51472.001-A awarded by the Space Telescope Science Institute, which is operated by the Association of Universities for Research in Astronomy, Incorporated, under NASA contract NAS5-26555. D.J.\ is supported by NRC Canada and by an NSERC Discovery Grant. The \textit{JWST} data presented in this paper were obtained from the Mikulski Archive for Space Telescopes (MAST) at the Space Telescope Science Institute. The specific observations analyzed can be accessed via \dataset[DOI: 10.17909/pmta-3239]{https://doi.org/10.17909/pmta-3239}.

\appendix

\section{Appendix A: IRTF Spectrum of MWC~758}
\label{Appendix A}
In this appendix we present the IRTF/SpeX spectrum with \textit{JWST}/NIRCam photometric filter bandpasses overlaid. 

\begin{figure*}[htpb]
\epsscale{1.15}
\plotone{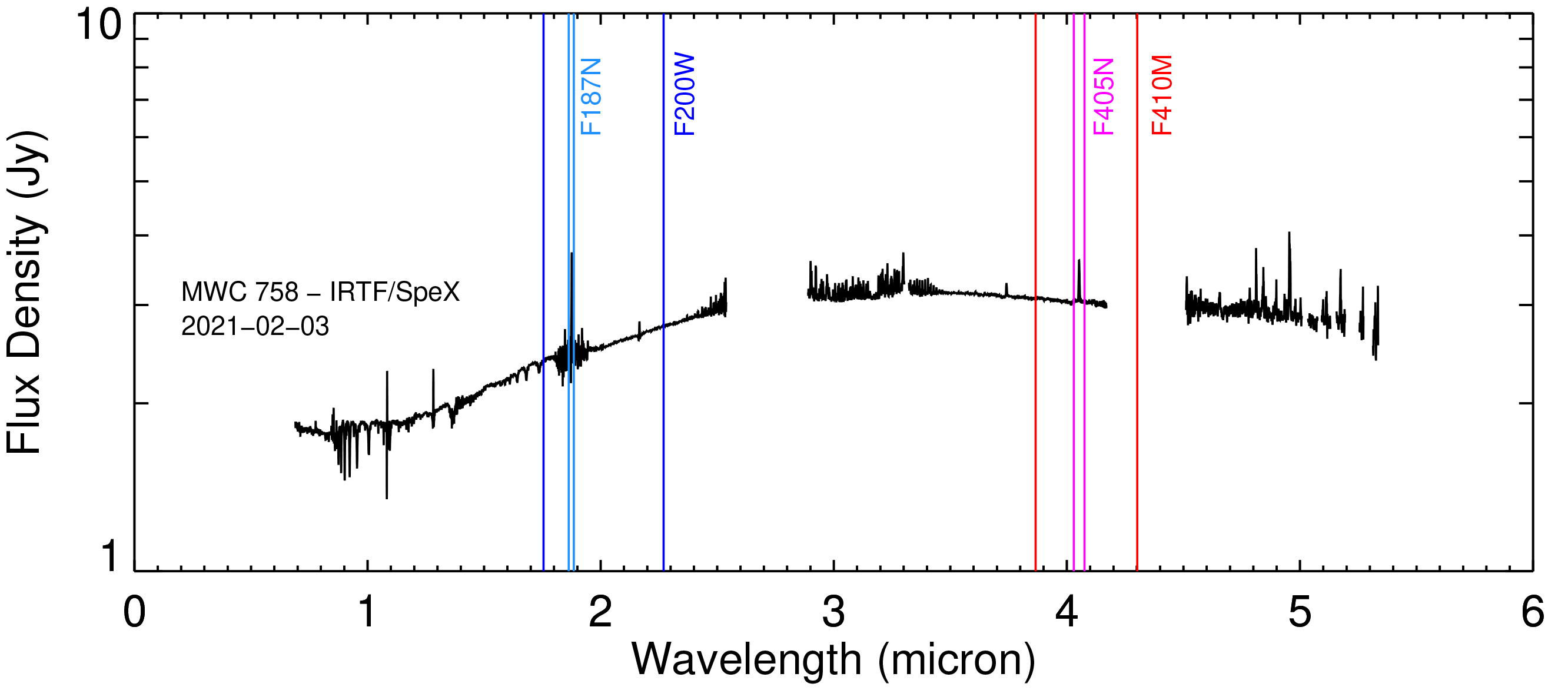}
\caption{IRTF/SpeX spectrum of MWC~758 taken on 2021-02-03. The flux of MWC~758 and its unresolved inner disk emission was used to convert from flux density to contrast units in Fig. \ref{cont}.}
\label{galfig}
\end{figure*}

\vspace{5cm}

\section{Appendix B: Wide-Field ADI Images}
\label{Appendix B}
In this appendix we show the wide-field ADI processed images. While there are several apparent point sources in the individual roll subtraction images, these are easily identifiable as speckles among the wings of MWC~758's PSF. No objects other than the known background star and newly discovered background galaxy are present in a stable position in both the Roll 1 - Roll 2 images and Roll 2 - Roll 1 images. 

\begin{figure*}[htpb]
\epsscale{1.1}
\plotone{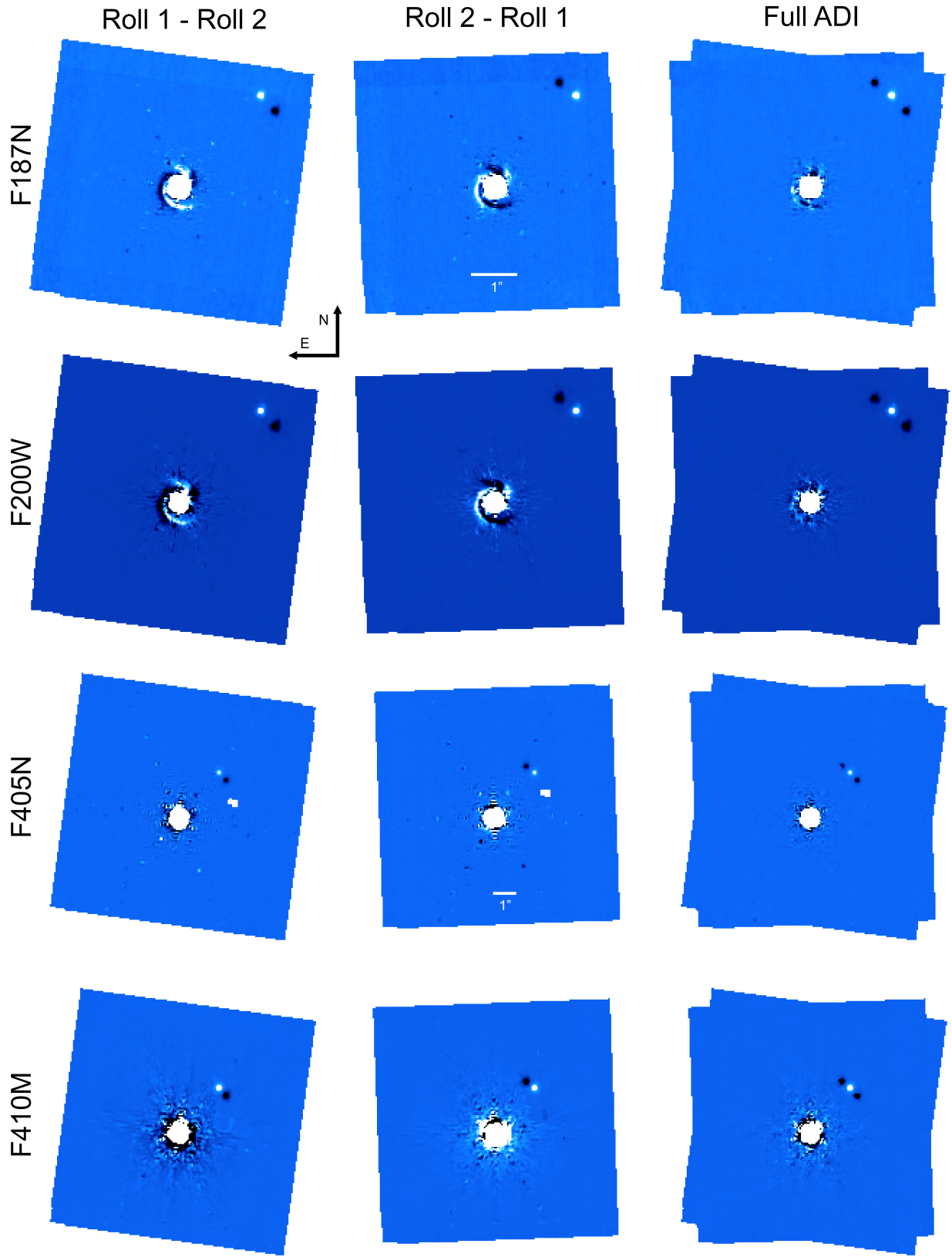}
\caption{Classical ADI images of each individual spacecraft roll and their combination. The images in the F187N and F200W filters (likewise F405N and F410M) share a common platescale. All images share a common orientation of North up and East left. The image stretch has been adjusted independently in order to show both the residual PSF noise and background noise, which varies depending on the wavelength and filter width. The background star is barely in the field of view in the short wavelength filters, and some detector edge effects are present near the bottom of the image. Positive and negative speckles are seen in some of the images resulting from wavefront drift of the primary mirror segments between the two roll orientations.}
\label{adi}
\end{figure*}
\clearpage

\section{Appendix C: Background Galaxy Image}
\label{Appendix C}
In this appendix we present the KLIP-processed F410M image showing a spatially elongated source $-$ identified as a background galaxy $-$ to the Northeast of MWC~758.

\begin{figure}[htpb]
\epsscale{0.5}
\plotone{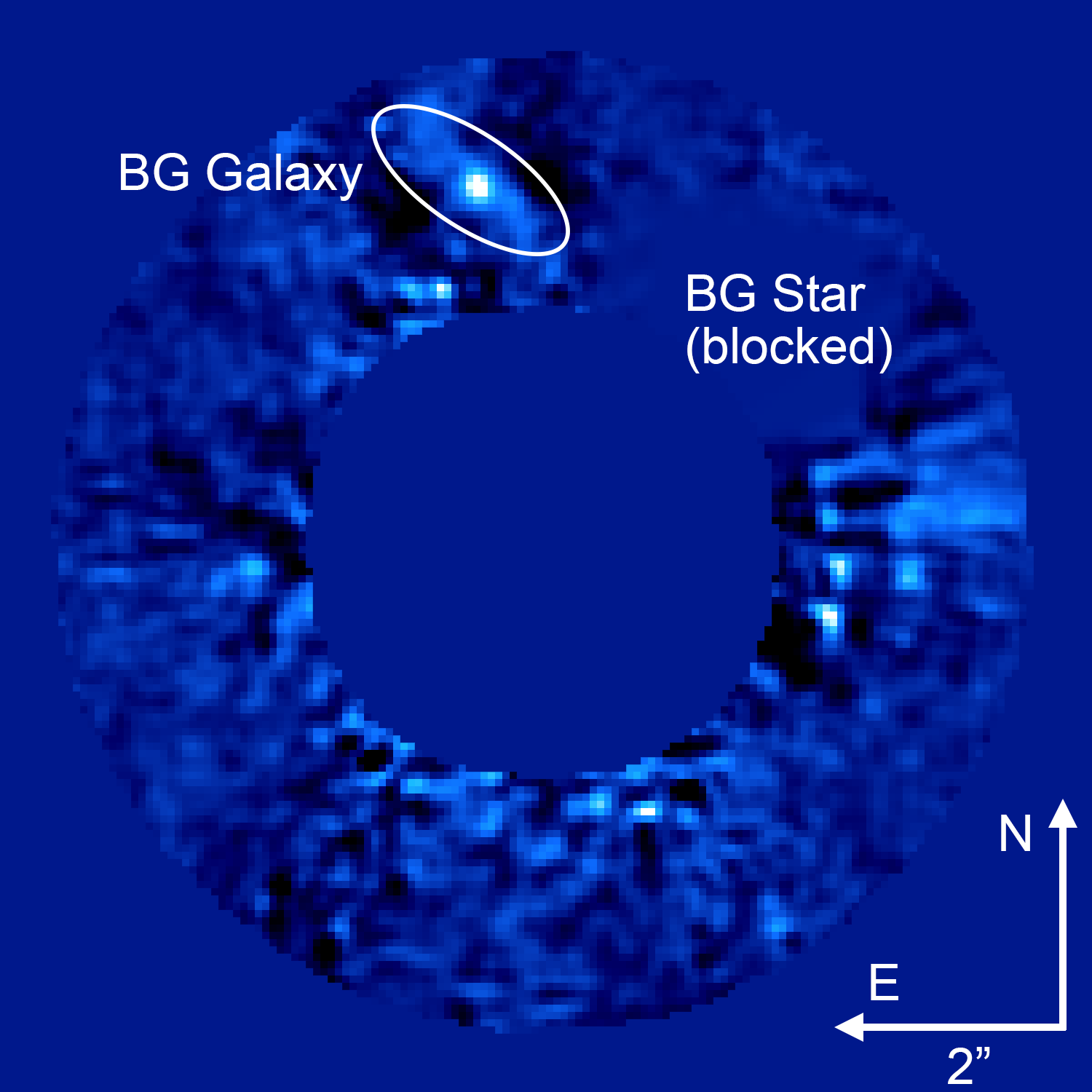}
\caption{Wide-field F410M image (processed with ADI-KLIP) with the central regions and those surrounding the known background star blocked in order to show the newly discovered faint background galaxy to the North of MWC758. The object's characterization arises from its spatially extended nature from the NE to SW. The galaxy is not detected in other filters.}
\label{galfig}
\end{figure}

\end{document}